\documentclass[12pt,a4paper]{article}

\usepackage[british]{babel}

\usepackage[a4paper,top=2cm,bottom=2cm,left=2.5cm,right=2.5cm,marginparwidth=1.75cm]{geometry}

\usepackage[style=apa, backend=biber]{biblatex} 
\DeclareLanguageMapping{british}{british-apa} 
\DeclareFieldFormat[article]{volume}{\apanum{#1}} 

\usepackage{amsmath}
\usepackage{graphicx}
\usepackage[colorlinks=true, allcolors=blue]{hyperref}
\usepackage{hyperref}
\usepackage[title]{appendix}
\usepackage{mathrsfs}
\usepackage{amsfonts}
\usepackage{booktabs} 
\usepackage{caption}  
\usepackage{threeparttable} 
\usepackage{algorithm}
\usepackage{algorithmicx}
\usepackage{algpseudocode}
\usepackage{listings}
\usepackage{enumitem}
\usepackage{chngcntr}
\usepackage{booktabs}
\usepackage{lipsum}
\usepackage{subcaption}
\usepackage{authblk}
\usepackage[T1]{fontenc}    
\usepackage{csquotes}       
\usepackage{diagbox}

\usepackage{setspace}
\usepackage{titlesec}
\titleformat{\section} 
  {\normalfont\Large\bfseries}{\thesection.}{1em}{}

\usepackage{float}   
\usepackage{caption} 


\title{Financial News-Driven LLM Reinforcement Learning for Portfolio Management}

\author{Ananya Unnikrishnan}

\date{November 17, 2024} 

\begin{document}

\maketitle

\begin{abstract}
Reinforcement learning (RL) has emerged as a transformative approach for financial trading, enabling dynamic strategy optimization in complex markets. This study explores the integration of sentiment analysis, derived from large language models (LLMs), into RL frameworks to enhance trading performance. Experiments were conducted on single-stock trading with Apple Inc. (AAPL) and portfolio trading with the ING Corporate Leaders Trust Series B (LEXCX). The sentiment-enhanced RL models demonstrated superior net worth and cumulative profit compared to RL models without sentiment and, in the portfolio experiment, outperformed the actual LEXCX portfolio's buy-and-hold strategy. These results highlight the potential of incorporating qualitative market signals to improve decision-making, bridging the gap between quantitative and qualitative approaches in financial trading.  
\end{abstract}

\section{Introduction}
\label{sec:introduction}
In recent years, reinforcement learning (RL) has gained traction in financial trading, providing algorithms that adapt and optimize trading strategies based on sequential decisions \parencite{jiang_deep_2017}. Unlike traditional rule-based systems, RL models can navigate complex market environments, continuously updating their strategies to maximize returns \parencite{deng_deep_2016}. However, despite their effectiveness in making data-driven decisions, RL algorithms often lack access to qualitative factors that influence market behavior—such as sentiment from news on stocks and the economy \parencite{bollen_twitter_2011}. This omission can limit their responsiveness to broader market moods and events that may not immediately reflect in stock prices but have significant impacts \parencite{zhang_enhancing_2023}.

Integrating sentiment analysis into RL for financial applications offers a way to bridge this gap. Large Language Models (LLMs), which excel in extracting sentiment from text, can transform qualitative insights from financial news into structured data for RL models to incorporate \parencite{yang_financial_2024}. This paper explores how sentiment analysis derived from LLMs can enhance the performance of RL algorithms, enabling them to capture market sentiment's immediate and lagged effects on stock prices and portfolios.

The objective of this research is to demonstrate that adding sentiment analysis to RL algorithms improves trading and portfolio management performance. Initially, we developed a baseline RL trading algorithm on a single stock (AAPL) to compare its performance with and without sentiment input. Following this, we extended the approach to a diversified portfolio based on stocks from the ING Corporate Leaders Trust Series B (LEXCX). By comparing the sentiment-enhanced RL model to both the standard RL model and the original LEXCX portfolio, we aim to show the potential of combining LLM-based sentiment analysis with reinforcement learning for portfolio optimization.

\section{Literature Review}
\label{sec:literature_review}
Reinforcement learning (RL) has shown promise in financial trading, offering adaptive algorithms that optimize trading actions in dynamic market conditions. Previous studies have demonstrated RL's effectiveness in single-stock trading, as well as in managing diversified portfolios by maximizing cumulative returns or minimizing risk over time. For example, several approaches, such as deep Q-learning and policy gradient methods, have been applied to simulate real-world trading scenarios, producing strategies that adapt to historical price patterns and technical indicators \parencite{yang_deep_2019}. Despite these advancements, RL in trading often relies solely on price and volume data, which may limit its ability to account for broader market signals and sentiment-driven shifts.

Sentiment analysis has increasingly become a tool for incorporating qualitative insights into financial models. By quantifying sentiment from news articles, social media, and analyst reports, models can enhance trading strategies with insights that capture public and institutional opinions. Research has shown that positive or negative sentiment can predict short-term price movements, while sustained sentiment trends often correlate with market volatility \parencite{li_sentiment_2021}. In portfolio management, sentiment analysis has been employed as an additional input to improve asset allocation and market timing decisions, suggesting its value in enhancing model performance \parencite{goud_integrating_2023}.

Recent advancements in large language models (LLMs), such as GPT and BERT, have enabled more accurate and nuanced sentiment analysis by extracting context-specific sentiment from financial text. These models have expanded sentiment analysis capabilities, allowing for more detailed and targeted sentiment scoring relevant to financial applications \parencite{qiu_large_2024}. Despite these advances, there is a gap in the literature regarding the integration of LLM-driven sentiment analysis into RL models specifically for portfolio management. This study addresses this gap by investigating how adding LLM-based sentiment analysis can improve RL model performance, particularly for portfolio strategies that benefit from qualitative insights alongside traditional quantitative data.

\section{Methodology}
\label{sec:methodology}

\subsection{Reinforcement Learning}
Reinforcement learning (RL) is a machine learning paradigm where an agent learns to make sequential decisions by interacting with an environment, aiming to maximize cumulative rewards. This interaction is often framed as a Markov Decision Process (MDP), defined by a tuple (S,A,P,R,$\gamma$):

\begin{itemize}
\item States (S): Represent different configurations of the environment that the agent observes at each time step.
\item Actions (A): Define possible decisions or actions that the agent can take to interact with the environment.
\item Transition Probability (P): Indicates the probability of transitioning from one state to another given a specific action.
\item Reward Function (R): Provides feedback on the outcome of actions, helping the agent learn what decisions yield high or low rewards.
\item Discount Factor ($\gamma$): Balances the importance of immediate rewards versus future rewards, allowing the agent to consider long-term benefits.
\end{itemize}

\begin{figure}[H]
 \centering
 \makebox[\textwidth][c]{\includegraphics[width=0.5\textwidth]{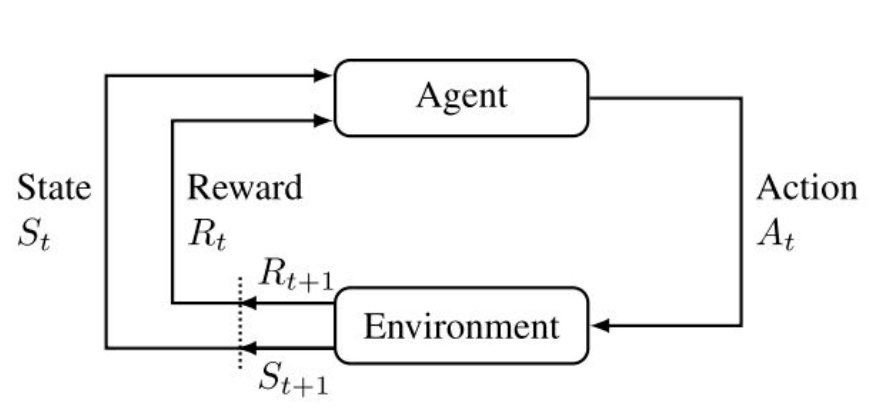}}
 \caption{MDP in Reinforcement Learning}
 \label{fig-1}
\end{figure}

The objective of an RL agent is to discover an optimal policy $\pi$, a mapping from states to actions that maximizes the expected cumulative reward over time \parencite{sutton_reinforcement_2018} as shown in Figure \ref{fig-1}.

Central to RL are value functions, which estimate the expected cumulative rewards associated with states or state-action pairs. These value functions help guide the agent in making decisions that maximize long-term rewards:

\textbf{State-Value Function} \( V^{\pi}(s) \): Calculates the expected return when starting from state \( s \) and following policy \( \pi \):
    \[
    V^{\pi}(s) = \mathbb{E}_{\pi} \left[ \sum_{t=0}^{\infty} \gamma^t R(s_t, a_t) \mid s_0 = s \right]
    \]

\textbf{Action-Value Function} \( Q^{\pi}(s, a) \): Represents the expected return of taking an action \( a \) in state \( s \) and subsequently following policy \( \pi \):
    \[
    Q^{\pi}(s, a) = \mathbb{E}_{\pi} \left[ \sum_{t=0}^{\infty} \gamma^t R(s_t, a_t) \mid s_0 = s, a_0 = a \right]
    \]

The optimal policy \( \pi^* \) is learned by maximizing these value functions. In practice, Q-learning and policy gradient methods are commonly employed to update and optimize the agent’s decision-making process \parencite{mnih_human_2015}.

\begin{figure}[H]
 \centering
 \makebox[\textwidth][c]{\includegraphics[width=0.5\textwidth]{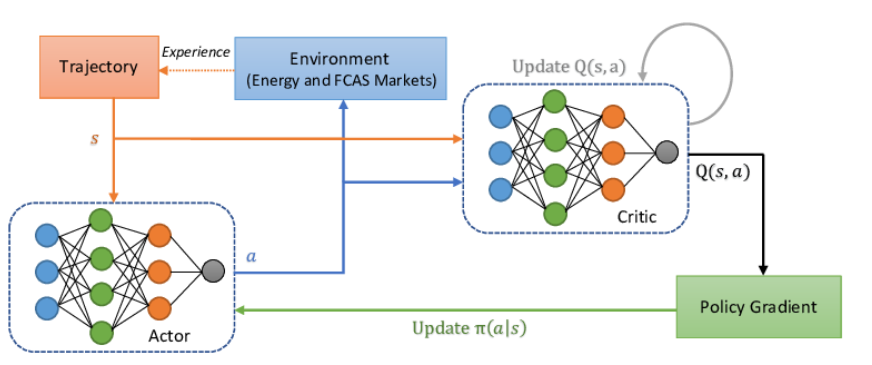}}
 \caption{Proximal Policy Optimization (PPO) architecture}
 \label{fig-2}
\end{figure}

In recent years, Deep Reinforcement Learning (DRL) has advanced RL applications to complex, high-dimensional environments by using neural networks to approximate value functions and policies. DRL algorithms, such as Deep Q-Networks (DQN) and Proximal Policy Optimization (PPO) as shown in Figure \ref{fig-2}, leverage deep learning architectures to make RL scalable and efficient in complex tasks. DQNs, for instance, approximate Q-values through neural networks, making them suitable for environments with large state and action spaces \parencite{mnih_human_2015}. PPO, a widely-used policy optimization algorithm, stabilizes learning for continuous action spaces by clipping the probability ratios, making it effective for dynamic and diverse decision-making tasks \parencite{schulman_proximal_2017}.

\subsection{Trading Reinforcement Learning Algorithm}
The trading reinforcement learning (RL) algorithm was structured to simulate trading decisions by balancing flexibility, reward optimization, and transaction cost considerations. Each aspect of the algorithm, from the action space to reward calculation, is designed to enable the agent to learn profitable trading behaviors while managing risks and avoiding excessive costs.

To build a trading reinforcement learning (RL) agent, we implemented a custom environment compatible with OpenAI Gym. This framework provides a standardized way to define RL environments, including observation spaces, action spaces, and step-based interactions. Our environment simulates the decision-making process by defining key elements: an action space that includes discrete and continuous actions for buying, selling, or holding positions, and an observation space capturing the state, which consists of relevant financial information at each time step \parencite{brockman_openai_2016}. This environment is structured to allow various RL algorithms, such as Proximal Policy Optimization (PPO), to interact with it and optimize a trading strategy over time.

The action space in this RL model is defined as a continuous, two-dimensional space allowing for nuanced trading decisions:
\begin{itemize}
\item Action Type: A scalar between 0 and 2, where values less than 1 represent a "buy" action, values between 1 and 2 represent a "sell" action, and values at 1 represent "hold."
\item Action Amount: A scalar between 0 and 0.5, indicating the proportion of the agent’s balance (for buying) or shares held (for selling) that will be traded.
\end{itemize}
This continuous space gives the agent precise control over trading sizes and allows it to adjust based on learned preferences and market conditions, promoting flexibility in trading. The continuous structure also better mimics real-world trading, where trades can vary in size rather than being restricted to fixed volumes \parencite{lillicrap_continuous_2015}.

When the agent selects an action, the algorithm interprets and scales the action based on the selected action type and amount:
\begin{itemize}
\item Buying: The algorithm calculates the maximum shares the agent can buy with its current balance, scaling the purchase based on the action amount.
\item Selling: If the action type represents a sell, the algorithm calculates the number of shares to sell based on the agent’s current holdings, again scaled by the action amount.
\end{itemize}
For example, if the agent’s action amount is 0.2, only 20 percent of the maximum possible shares are bought or sold. This scaling mechanism allows the agent to adjust its market exposure without fully committing to a single decision, promoting a balanced trading approach. This dynamic scaling of trades helps prevent excessive risks and aligns with the agent’s goal to learn strategies that balance profitability and stability.

The reward function in this algorithm is designed to encourage both profitability and stability. The reward at each step includes the following components:
\begin{itemize}
\item Profit-Based Reward: The primary reward is based on the agent’s net change in account balance, accounting for both held shares (unrealized gains) and cash from sold shares (realized gains).
\item Balance Stability Penalty: To discourage large swings in balance, a stability penalty is applied based on the deviation from an initial balance, rewarding more stable account values and promoting less volatile behavior.
\item Transaction Cost Penalty: A penalty proportional to the trade amount simulates real-world transaction fees, incentivizing the agent to minimize unnecessary trades.
\end{itemize}
The final reward equation combines these elements to balance profit-seeking, stability, and cost minimization. This reward function structure helps train the agent to make informed, responsible decisions, promoting growth in net worth without over-trading \parencite{schulman_proximal_2017}.

\subsection{Integration of Sentiment Analysis}
Incorporating sentiment analysis into the RL algorithm allows the trading agent to consider market sentiment, adding a qualitative dimension to its decision-making process. Sentiment data, extracted from financial news, is mapped to quantitative values and integrated into the algorithm’s state observations, action adjustments, and reward calculations. This additional context helps the agent align its trading strategy with prevailing market sentiment, which can be a powerful indicator of price movements \parencite{chen_wisdom_2019}.

Sentiment data is included in the agent’s observation space, providing insights into market sentiment alongside traditional financial metrics like price and volume. The sentiment data is scaled to a range of [-1,1], where positive values represent optimistic sentiment (e.g., “Positive” or “Extremely Positive”) and negative values represent pessimistic sentiment (e.g., “Negative” or “Extremely Negative”) and zero represents Neutral. By appending this scaled sentiment value to each observation, the agent can consider sentiment trends as part of its decision-making process. This integration effectively extends the observation space to include sentiment, allowing the agent to detect shifts in sentiment that may precede price changes.

To align trading actions with sentiment, the sentiment score influences the action amount taken by the agent. For example: When the sentiment is positive, the agent’s buy amount is slightly increased, encouraging it to capitalize on positive market momentum. Conversely, when the sentiment is negative, the agent’s sell amount is increased, helping it reduce exposure in potentially unfavorable conditions. This adjustment is implemented by adding or subtracting a factor (0.1 times the sentiment score) to the action amount. Positive sentiment thus leads to a higher buy percentage, while negative sentiment leads to a higher sell percentage. This sentiment-based bias makes the agent more responsive to market sentiment, effectively enabling it to “lean” in the direction suggested by recent news sentiment \parencite{bollen_twitter_2011}.

The reward function is also adjusted to incorporate sentiment alignment. Beyond traditional metrics like net worth growth and balance stability, the reward function includes a sentiment alignment reward:
\begin{itemize}
\item Sentiment Reward: If the sentiment score aligns with the price movement direction (e.g., positive sentiment and rising prices), the agent receives an additional reward proportional to the sentiment score. This reward encourages the agent to make trades that align with market sentiment.
\item Volatility Adjustment: To prevent the agent from relying solely on sentiment in volatile conditions, this reward is scaled by recent price volatility. In high-volatility periods, the sentiment reward is reduced slightly to account for the unpredictability of price movements. The reward function thus incentivizes the agent to integrate sentiment effectively, while adjusting its reliability based on recent market stability.
\end{itemize}
This sentiment-based reward mechanism encourages the agent to make sentiment-aligned trades when recent sentiment is predictive of price direction. The additional reward terms guide the agent in optimizing both profitability and sentiment sensitivity, enhancing its performance in sentiment-driven markets.

\subsection{Extension to Portfolio Management}
Expanding the reinforcement learning (RL) algorithm to support portfolio management introduces several new complexities. In this approach, the agent must manage a collection of assets, making trading decisions that consider individual asset sentiments and market conditions. Each asset has its own price movements and sentiment data, and the agent must learn how to allocate capital across assets effectively to maximize the portfolio’s overall net worth. The portfolio trading environment is designed to integrate both stock-specific features and sentiment data to enable more sophisticated, sentiment-aware portfolio management.

In a portfolio setting, each asset (stock) in the environment has its own observation space, which includes price data, account information, and sentiment metrics. The observation space now consists of a matrix, where each row corresponds to a different stock. Key features in each row include:
\begin{itemize}
\item Price Data: The last five time steps of Open, High, Low, Close, and Volume data, normalized to provide a consistent input range.
\item Account Information: The agent’s current holdings, balance, and cost basis for each stock.
\item Sentiment Data: A scaled sentiment score for each stock, mapped to the range [-1,1] for consistency with other normalized values.
\end{itemize}
By expanding the observation space to include sentiment data for each stock, the agent gains insight into how market sentiment varies across its portfolio. This multi-asset setup provides the agent with a broader context, enabling it to make sentiment-driven decisions for each individual stock \parencite{chen_wisdom_2019}.

The action space in the portfolio environment allows the agent to take independent actions for each asset in the portfolio. This means that for a portfolio with n assets, the action space consists of n continuous values, each representing a decision for a specific asset. Each action can range from 0 to 2, where:
\begin{itemize}
\item Values below 1: Indicate a “buy” action for the stock.
\item Values between 1 and 2: Indicate a “sell” action.
\end{itemize}

Sentiment data for each stock influences the action amount. For example: When the sentiment score is positive, the action amount is adjusted to increase the buy amount, encouraging the agent to invest more in stocks with positive sentiment. Conversely, when sentiment is negative, the action amount is scaled to increase sell pressure, reducing exposure in stocks with negative sentiment.
This sentiment-adjusted action mechanism allows the agent to react more dynamically to sentiment changes, enhancing its ability to allocate resources across the portfolio based on sentiment cues \parencite{bollen_twitter_2011}.

The reward function in the portfolio environment is designed to optimize the entire portfolio’s net worth, with adjustments for sentiment alignment and transaction costs. The reward calculation includes:
\begin{itemize}
\item Portfolio Net Worth Change: The primary reward component is based on changes in the portfolio’s total net worth, calculated by summing the value of all stocks held and cash balance.
\item Sentiment Alignment Reward: A sentiment-based reward is added when the agent’s trades align with both sentiment and price trends. For example, if sentiment is positive and the price increases, the agent receives a positive reward for aligning with sentiment. This reward reinforces sentiment-driven decisions that are supported by actual price movements.
\item Volatility Adjustment: The sentiment reward is scaled by recent price volatility for each stock, reducing the weight of sentiment in highly volatile conditions. This adjustment helps prevent the agent from over-relying on sentiment in unstable markets.
\end{itemize}
This multi-part reward structure encourages the agent to maximize net worth while considering sentiment-based insights and managing transaction costs. The sentiment alignment reward encourages trades that align with prevailing market sentiment, improving the agent’s overall responsiveness to sentiment \parencite{schulman_proximal_2017}.

\section{Experiments and Results}
\label{sec:results}

\subsection{Data Pre-Processing}
Data pre-processing for our model comprises gathering both quantitative stock market data and qualitative sentiment data, ensuring that the reinforcement learning (RL) model benefits from a comprehensive view of market dynamics. This process was applied to both single-stock trading (using Apple Inc., AAPL) and portfolio trading (ING Corporate Leaders Trust Series B).

For the single-stock trading model, we obtained historical daily trading data for Apple Inc. (AAPL), capturing Open, High, Low, Close, and Volume prices. Similarly, for portfolio trading, we collected data for each stock in the ING Corporate Leaders Trust Series B (LEXCX) portfolio. The stocks included in this portfolio are:
\begin{itemize}
\item Union Pacific Corp. (UNP)
\item Berkshire Hathaway Inc. (BRK-B)
\item Marathon Petroleum Corp. (MPC)
\item Exxon Mobil Corp. (XOM)
\item Linde plc (LIN)
\item Procter \&\ Gamble Co. (PG)
\item Chevron Corp. (CVX)
\item Comcast Corp. (CMCSA)
\item Praxair Inc. (PX)
\item ConocoPhillips (COP)
\item General Electric Co. (GE)
\item Honeywell International Inc. (HON)
\item International Paper Co. (IP)
\item Johnson \&\ Johnson (JNJ)
\item Merck \&\ Co. (MRK)
\item Pfizer Inc. (PFE)
\item Raytheon Technologies Corp. (RTX)
\item Texas Industries Inc. (TX)
\item Wells Fargo \&\ Co. (WFC)
\item Weyerhaeuser Co. (WY)
\item DuPont de Nemours Inc. (DD)
\end{itemize}

Using the Yahoo Finance API, we gathered data spanning from November 16, 2023, to November 10, 2024. This data provides essential quantitative features, enabling the RL agent to detect price trends and trading volumes. For ease of analysis and integration, each stock's data was saved as a separate CSV file.

To introduce qualitative information into the model, we employed sentiment analysis by retrieving daily news articles for each stock from the Finnhub API. For each trading day within the specified range, we extracted and aggregated article headlines and summaries related to AAPL (for single-stock trading) and each symbol in the LEXCX portfolio (for portfolio trading). We then used OpenAI’s GPT-based large language model (LLM) to generate a daily sentiment score for each stock. The LLM was prompted to classify the sentiment of each day's news coverage as one of five categories: 
\begin{itemize}
\item Extremely Negative
\item Negative
\item Neutral
\item Positive
\item Extremely Positive 
\end{itemize}
This categorization captures daily sentiment trends, providing the RL agent with a qualitative insight into market mood and anticipated price direction. An adaptive retry mechanism was used to handle potential API rate limits or network interruptions, ensuring that data for the entire period was successfully collected.

For both AAPL and the LEXCX portfolio stocks, price data and daily sentiment scores were merged on a date-by-date basis. Missing sentiment values were filled with a neutral score to maintain data continuity. This combined dataset serves as the input for our RL model, enhancing it with both quantitative price data and qualitative sentiment information. As a result, the RL agent can adapt its trading strategies in response to both price patterns and sentiment-driven shifts, potentially improving model responsiveness to market events and trends.

\subsection{Experiments}
The experimental design evaluates the performance of a reinforcement learning (RL) agent in a simulated stock trading environment, with two setups tested: a single-stock trading model focused on Apple Inc. (AAPL) and a portfolio-based trading model incorporating sentiment analysis for the ING Corporate Leaders Trust Series B (LEXCX). For each experiment, the model's success was measured by net worth, balance, and cumulative profit over multiple episodes, both with and without sentiment integration. Additionally, the portfolio experiment included a baseline comparison against the actual LEXCX portfolio performance for the same period, using the same initial investment amount. This allowed us to assess the overall effectiveness of the RL algorithm relative to a standard buy-and-hold approach.

\subsubsection{Single-Stock Experiment}
In the single-stock experiment, we developed a custom environment using OpenAI Gym to simulate trading conditions for Apple Inc. (AAPL) over the historical period from our dataset. The agent's objective was to maximize its net worth by making sequential buy, hold, or sell decisions based on observed stock data and account information.

\textbf{Environment Setup}: The environment was initialized with daily trading data for AAPL, covering essential stock features such as Open, High, Low, Close, and Volume. Additional account-related information, including balance, cost basis, and net worth, was also included in the observation space. The environment's action space was defined as a two-dimensional continuous space, allowing the agent to decide on both the type of action (buy/sell/hold) and the amount to trade.

\textbf{Observation Space and Action Execution}: The observation space consisted of the stock's recent five-day data points alongside the agent's balance, net worth, and cost basis. To optimize trading strategy, the agent's actions were interpreted as either a buy or sell operation, with trade size determined by the action amount. This enabled flexibility in trade size, allowing the agent to scale its investment based on its confidence in the stock’s direction.

\textbf{Reward Calculation}: The reward function was designed to encourage both profitability and stability. It factored in net worth growth, penalizing excessive fluctuations in balance to promote steady gains. Each transaction incurred a small penalty proportional to trade size, representing real-world trading costs, thereby discouraging high-frequency trades. Episodes concluded if the agent’s net worth dropped to zero or after a fixed number of steps.

\textbf{Training and Performance Metrics}: The model was trained using Proximal Policy Optimization (PPO) for 20,000 timesteps. Following training, the model was evaluated over 100 episodes, each running for 2,000 timesteps, to assess its average performance. Key metrics recorded at each episode’s end included final net worth, remaining balance, and cumulative profit. The average performance across episodes was calculated to gauge the model's robustness in single-stock trading.

\subsubsection{Portfolio Experiment}
In the portfolio experiment, we extended the single-stock trading approach to a diversified portfolio, focusing on the ING Corporate Leaders Trust Series B (LEXCX) and integrating sentiment analysis for each stock. The objective was to assess the RL agent’s performance in actively managing a multi-asset portfolio compared to a passive investment benchmark.

\textbf{Environment Setup}: Building on the single-stock setup, the portfolio environment was adapted to allow the RL agent to make independent Buy, Hold, or Sell decisions for each stock in the LEXCX portfolio. In the sentiment-enhanced environment, daily sentiment scores were incorporated, allowing the agent to adjust its actions based on qualitative market insights. This extension to a multi-stock setting required the observation space to scale, capturing each stock’s recent five-day trading data and account metrics specific to each stock, such as balance and net worth contributions.

\textbf{Observation Space and Action Execution}: The RL agent’s observation space included price data and sentiment information for all stocks, enabling it to make decisions based on both quantitative and qualitative data. By structuring the action space to accommodate each stock, the agent could buy, hold, or sell any combination of stocks in the portfolio, with action amounts scaled by sentiment data when available. This flexibility allowed the agent to dynamically adjust holdings and diversify its positions based on the market environment and sentiment shifts.

\textbf{Reward Calculation}: The reward function encouraged consistent portfolio growth and penalized high trading frequency to promote steady returns. For the sentiment-enhanced environment, an additional reward adjustment based on alignment with sentiment provided positive reinforcement when sentiment-driven decisions aligned with actual price trends. This setup enabled the RL agent to leverage sentiment data without over-relying on it, balancing both stability and adaptability in its trading strategy. Episodes ended when the agent’s total net worth fell below a critical threshold or reached a fixed step limit.

\textbf{Baseline Comparison}: To establish a benchmark, we calculated the net worth and cumulative profit for the actual performance of the LEXCX portfolio over the same period. We used an initial investment of \$10,000, assuming this amount was invested directly in the LEXCX fund and held without any additional trades. Historical price data for LEXCX provided the basis for tracking the portfolio’s net worth and cumulative profit over the investment period. This comparison allowed us to evaluate the RL agent’s ability to outperform the real-world performance of the LEXCX portfolio, highlighting the potential added value of the RL agent’s active trading decisions and sentiment-driven adjustments relative to a standard buy-and-hold approach within an established fund.

\textbf{Training and Performance Metrics}: The RL model was trained using Proximal Policy Optimization (PPO) for 20,000 timesteps and evaluated over 100 episodes, each with 2,000 timesteps. Key metrics collected included net worth, balance, and cumulative profit at each episode’s conclusion. Average performance across episodes demonstrated the RL agent’s ability to adapt its strategy in response to sentiment data, highlighting the potential added value of sentiment-aware trading. By contrasting these results with the actual LEXCX portfolio’s net worth and profit over the same period, we assessed the effectiveness of the RL agent’s approach to dynamic portfolio management relative to the established fund's performance.

\subsection{Results}
The results of the experiments are presented in two sections: first, the single-stock trading experiment (AAPL), and second, the portfolio trading experiment (ING Corporate Leaders Trust Series B). Each section includes analyses of the RL model without sentiment integration and the sentiment-enhanced RL model. Key metrics, including net worth, cumulative profit, and balance, are evaluated across episodes and timesteps. For the portfolio experiments, the RL model’s performance is further compared to the actual performance of the LEXCX portfolio over the same time frame, providing insights into the effectiveness of active trading strategies. The visualizations highlight the model's ability to adapt to market dynamics and leverage sentiment data to improve decision-making.

\subsubsection{Single-Stock Experiment Results}

\begin{figure}[H]
 \centering
 \makebox[\textwidth][c]{\includegraphics[width=0.8\textwidth]{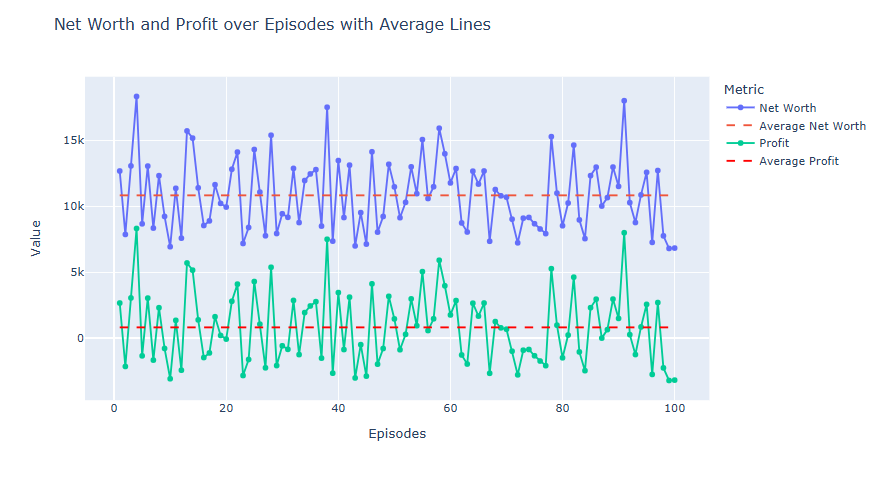}}
 \caption{Net Worth and Profit Over Episodes Without Sentiment Analysis}
 \label{fig-3}
\end{figure}

The performance of the RL model without sentiment analysis is illustrated in Figure \ref{fig-3}. Over 100 evaluation episodes, the agent achieved an average net worth of \$10,825.41, while the average profit remained at \$825.41. These results demonstrate the RL agent's ability to effectively navigate single-stock trading strategies for Apple Inc. (AAPL) based solely on quantitative data.

\begin{figure}[H]
 \centering
 \makebox[\textwidth][c]{\includegraphics[width=0.8\textwidth]{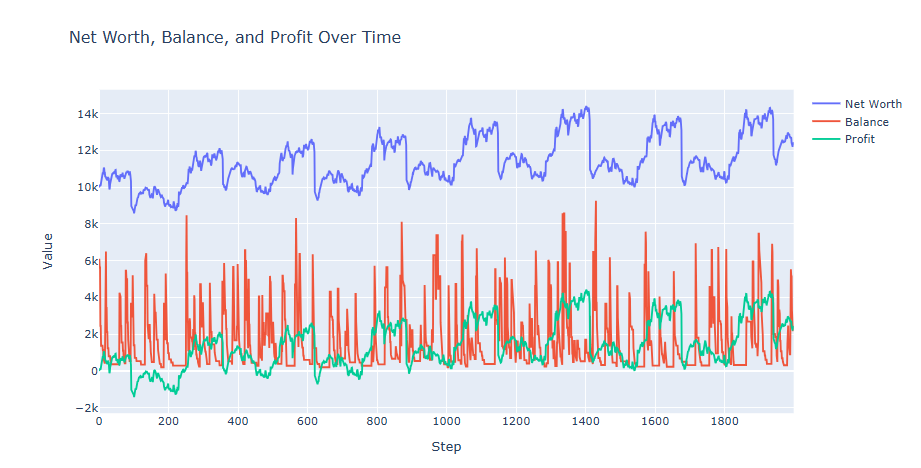}}
 \caption{Net Worth, Balance, and Profit Over Time Without Sentiment Analysis}
 \label{fig-4}
\end{figure}

Figure \ref{fig-4} shows the evolution of net worth, balance, and profit over a single episode for the RL model without sentiment analysis. The agent demonstrated a consistent ability to grow net worth while maintaining a stable balance, reflecting its strategic use of available capital to achieve steady gains.

\begin{figure}[H]
 \centering
 \makebox[\textwidth][c]{\includegraphics[width=0.8\textwidth]{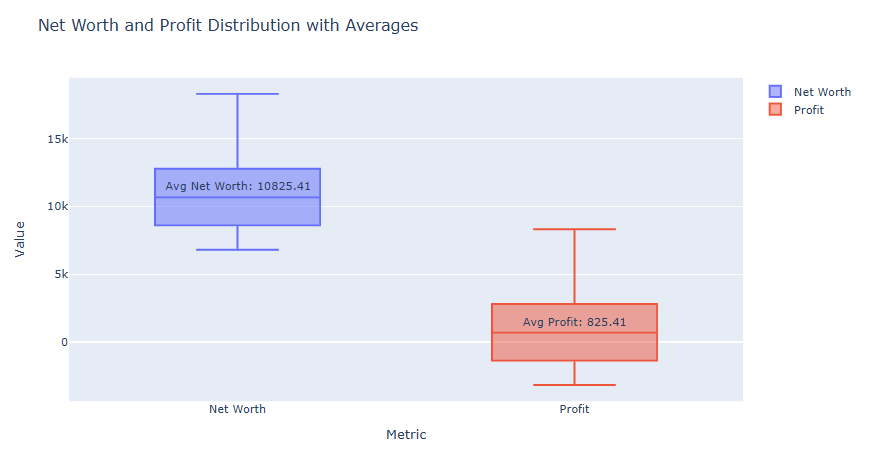}}
 \caption{Net Worth and Profit Distribution Without Sentiment Analysis}
 \label{fig-5}
\end{figure}

Figure \ref{fig-5} presents a box plot showing the distribution of net worth and profit across all episodes for the single-stock RL agent. The narrow range of values indicates reliable performance, with minimal variability between episodes, demonstrating the agent's robust decision-making capabilities in this controlled environment.

\begin{figure}[H]
 \centering
 \makebox[\textwidth][c]{\includegraphics[width=0.8\textwidth]{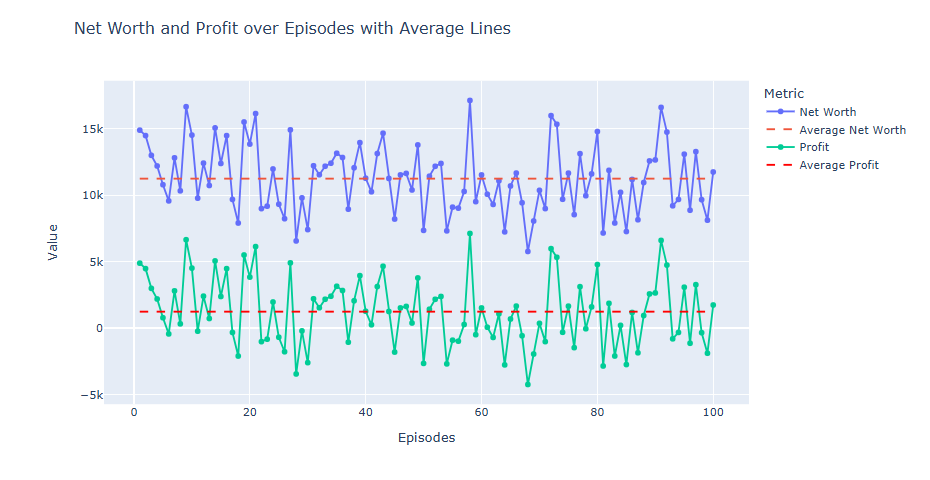}}
 \caption{Net Worth and Profit Over Episodes With Sentiment Analysis}
 \label{fig-6}
\end{figure}

Figure \ref{fig-6} illustrates the RL model’s performance when enhanced with sentiment analysis for AAPL. Incorporating sentiment data allowed the agent to achieve an average net worth of \$11,259.51, with an average profit of \$1,259.51. These results underscore the agent's ability to utilize qualitative data to make more informed trading decisions, resulting in improved outcomes compared to the baseline model.

\begin{figure}[H]
 \centering
 \makebox[\textwidth][c]{\includegraphics[width=0.8\textwidth]{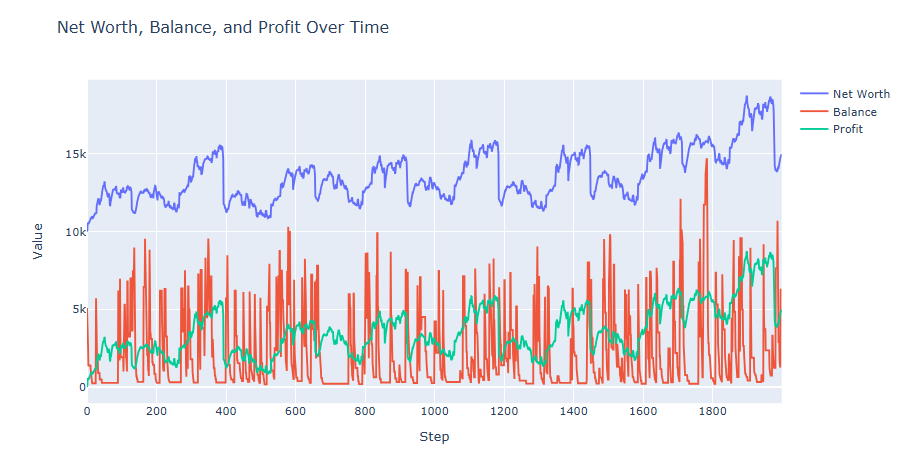}}
 \caption{Net Worth, Balance, and Profit Over Time With Sentiment Analysis}
 \label{fig-7}
\end{figure}

Figure \ref{fig-7} displays the evolution of net worth, balance, and profit over a single episode with sentiment integration. The sentiment-enhanced RL agent exhibited improved performance by effectively aligning its trades with market sentiment, leading to higher cumulative gains over the episode.

\begin{figure}[H]
 \centering
 \makebox[\textwidth][c]{\includegraphics[width=0.8\textwidth]{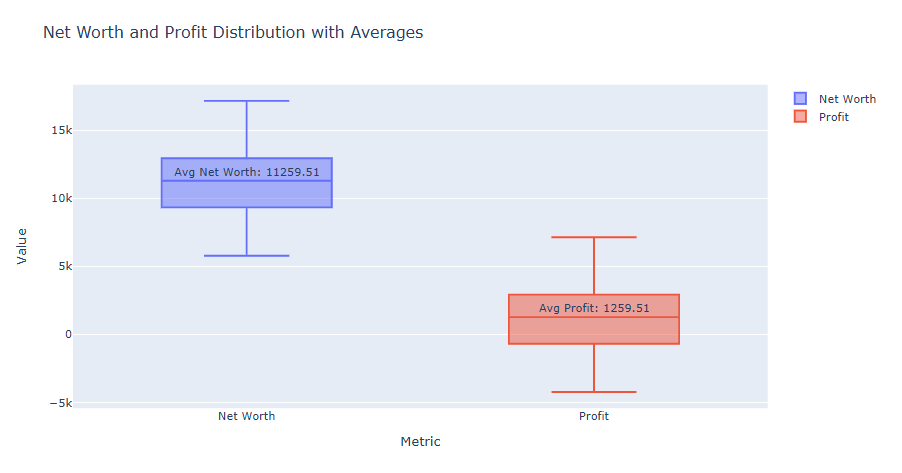}}
 \caption{Net Worth and Profit Distribution With Sentiment Analysis}
 \label{fig-8}
\end{figure}

Figure \ref{fig-8} presents the distribution of net worth and profit across evaluation episodes for the sentiment-enhanced RL agent. The results indicate a noticeable improvement in median values compared to the baseline model, highlighting the added value of incorporating sentiment data into the trading strategy.

\begin{table}[H]
\centering
\begin{tabular}{||c c c||} 
 \hline
  & RL (No Sentiment) & RL with Sentiment\\ [0.5ex] 
 \hline\hline
 Average Profit & \$825.41 & \$1,259.5\\ 
 \hline
 Average Net Worth & \$10,825.41 & \$11,259.51\\
 \hline
\end{tabular}
\caption{Comparison of Average Profit and Net Worth Across Single-Stock Experiments}
\label{tab-I}
\end{table}

Table \ref{tab-I} provides a summary of the average profit and net worth achieved across the three scenarios: RL without sentiment, RL with sentiment, and the actual LEXCX portfolio. The results demonstrate the RL agent’s superior performance, particularly when sentiment data was integrated, emphasizing its potential for dynamic portfolio management.

\subsubsection{Portfolio Experiment Results}

\begin{figure}[H]
 \centering
 \makebox[\textwidth][c]{\includegraphics[width=0.8\textwidth]{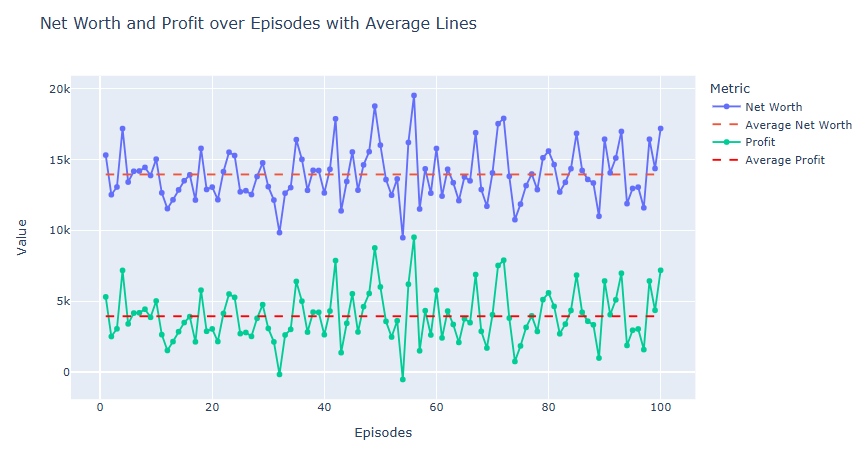}}
 \caption{Net Worth and Profit Over Episodes Without Sentiment Analysis}
 \label{fig-9}
\end{figure}

The performance of the RL model without sentiment analysis is illustrated in Figure \ref{fig-9}. Over 100 evaluation episodes, the agent achieved an average net worth of \$13,952.29, while the average profit remained at \$3,952.29. These results reflect consistent returns achieved through the RL agent's active trading strategy.

\begin{figure}[H]
 \centering
 \makebox[\textwidth][c]{\includegraphics[width=0.8\textwidth]{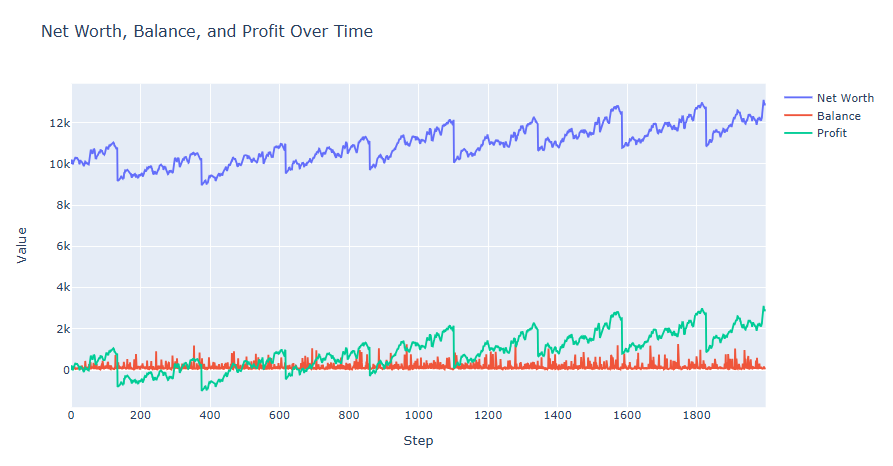}}
 \caption{Net Worth, Balance, and Profit Over Time Without Sentiment Analysis}
 \label{fig-10}
\end{figure}

Figure \ref{fig-10} demonstrates the evolution of the RL agent’s portfolio over individual steps in a single episode. The agent consistently increased its net worth while maintaining a stable balance, ensuring steady cumulative profit over time.

\begin{figure}[H]
 \centering
 \makebox[\textwidth][c]{\includegraphics[width=0.8\textwidth]{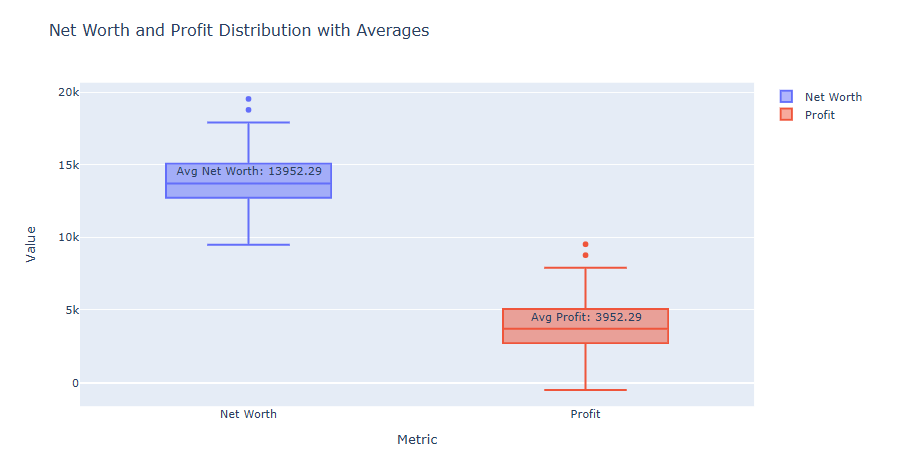}}
 \caption{Net Worth and Profit Distribution Without Sentiment Analysis}
 \label{fig-11}
\end{figure}

Figure \ref{fig-11} presents a box plot of the distribution of net worth and profit across all evaluation episodes. The median values closely align with the averages, reflecting stability in the agent’s performance despite minor variability across episodes.

\begin{figure}[H]
 \centering
 \makebox[\textwidth][c]{\includegraphics[width=0.8\textwidth]{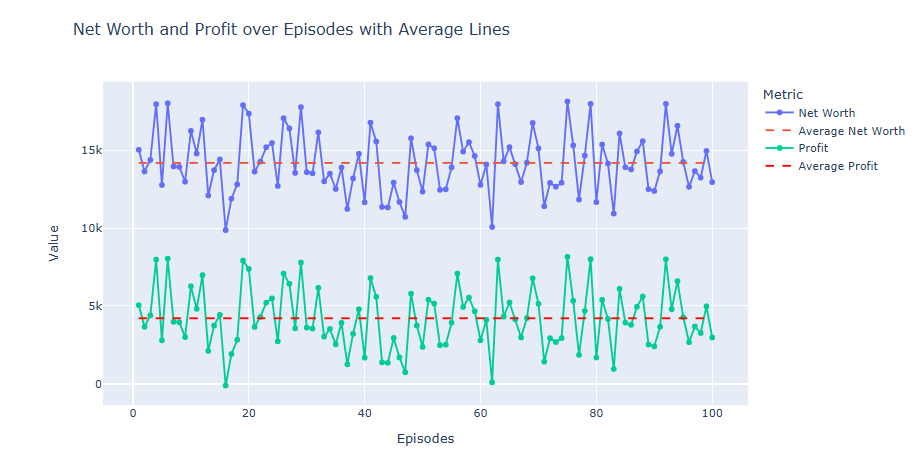}}
 \caption{Net Worth and Profit Over Episodes With Sentiment Analysis}
 \label{fig-12}
\end{figure}

Figure \ref{fig-12} illustrates the RL model’s performance when sentiment data was incorporated. The sentiment-enhanced RL model achieved an average net worth of \$14,201.94 and an average profit of \$4,201.94. This increase indicates the model's ability to leverage qualitative data to improve decision-making.

\begin{figure}[H]
 \centering
 \makebox[\textwidth][c]{\includegraphics[width=0.8\textwidth]{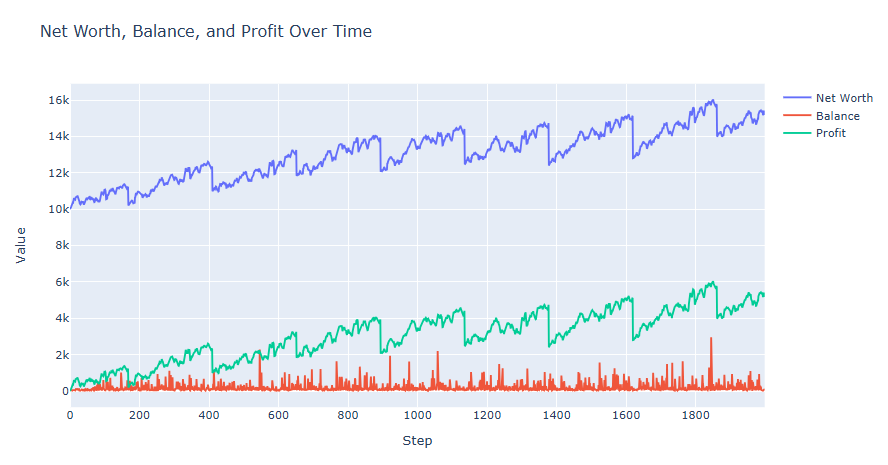}}
 \caption{Net Worth, Balance, and Profit Over Time With Sentiment Analysis}
 \label{fig-13}
\end{figure}

Figure \ref{fig-13} shows the evolution of net worth, balance, and profit over a single episode for the sentiment-enhanced RL model. The agent demonstrated a stronger ability to capitalize on profitable trades compared to the baseline RL model, resulting in higher overall gains.

\begin{figure}[H]
 \centering
 \makebox[\textwidth][c]{\includegraphics[width=0.8\textwidth]{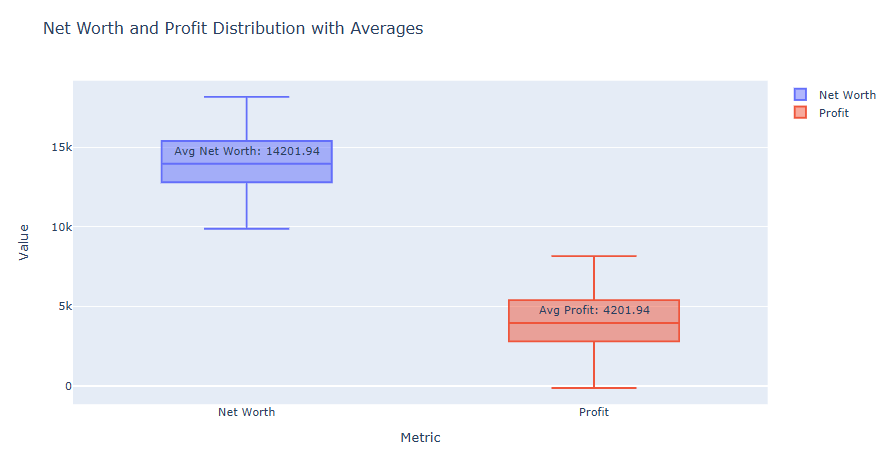}}
 \caption{Net Worth and Profit Distribution With Sentiment Analysis}
 \label{fig-14}
\end{figure}

Figure \ref{fig-14} displays the distribution of net worth and profit across episodes for the sentiment-enhanced RL model. The results show a higher median net worth and profit compared to the baseline RL model, emphasizing the value of incorporating sentiment data.

\begin{figure}[H]
 \centering
 \makebox[\textwidth][c]{\includegraphics[width=0.8\textwidth]{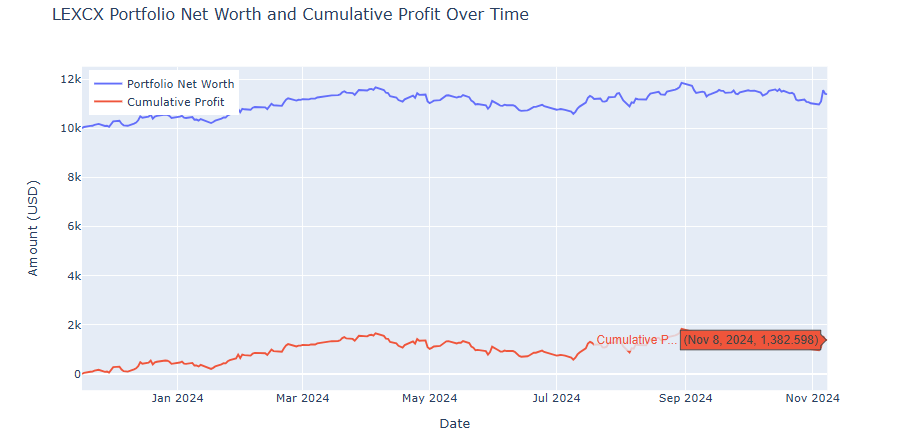}}
 \caption{Net Worth and Profit of the Actual LEXCX Portfolio}
 \label{fig-15}
\end{figure}

To benchmark the RL agent’s performance, we calculated the net worth and cumulative profit for a \$10,000 initial investment directly in the actual LEXCX portfolio, as shown in Figure \ref{fig-15}. The portfolio achieved a final net worth of \$11,382.60 and a cumulative profit of \$1,382.60 over the investment period. These results highlight the added value of the RL agent’s active trading strategy, particularly the sentiment-enhanced RL model, which consistently outperformed the real-world performance of the LEXCX portfolio.

\begin{table}[H]
\centering
\begin{tabular}{||c c c c||} 
 \hline
  & RL (No Sentiment) & RL with Sentiment & Actual LEXCX\\ [0.5ex] 
 \hline\hline
 Average Profit & \$3,952.29 & \$4,201.94 & \$1,382.60 \\ 
 \hline
 Average Net Worth & \$13,952.29 & \$14,201.94 & \$11,382.60 \\
 \hline
\end{tabular}
\caption{Comparison of Average Profit and Net Worth Across Portfolio Experiments}
\label{tab-II}
\end{table}

Table \ref{tab-II} provides a summary of the average profit and net worth achieved across the three scenarios: RL without sentiment, RL with sentiment, and the actual LEXCX portfolio. The results demonstrate the RL agent’s superior performance, particularly when sentiment data was integrated, emphasizing its potential for dynamic portfolio management.

\section{Discussion}
\label{sec:discussion}
The results of the experiments underscore the effectiveness of reinforcement learning (RL) in financial trading, particularly when enhanced with sentiment analysis. In the single-stock experiment with Apple Inc. (AAPL), the RL agent demonstrated a consistent ability to navigate market dynamics, achieving notable gains both with and without sentiment integration. The incorporation of sentiment analysis further improved trading outcomes, increasing the average net worth and profit across evaluation episodes. This suggests that qualitative data, such as sentiment extracted from financial news, can provide the RL agent with valuable contextual insights, allowing for more informed and adaptive decision-making.

In the portfolio experiment, the RL agent similarly outperformed the actual LEXCX portfolio, highlighting the potential advantages of active trading over passive investment strategies. The sentiment-enhanced RL model consistently delivered higher returns than both the RL model without sentiment and the LEXCX portfolio’s buy-and-hold performance. This demonstrates that the integration of sentiment data can improve the agent’s ability to capitalize on market trends and adapt to changing conditions. By leveraging both quantitative and qualitative data, the sentiment-enhanced RL model showcased its potential to generate higher cumulative profits and net worth, reinforcing the role of hybrid approaches in portfolio optimization.

The addition of sentiment analysis to the RL models aligns with previous research suggesting that market sentiment influences asset prices and volatility. By incorporating sentiment data, the RL agent was able to account for broader market dynamics that may not be immediately reflected in stock price movements. This approach allowed the sentiment-enhanced model to achieve more robust performance, particularly in the portfolio experiment, where it outperformed both the baseline RL model and the LEXCX portfolio.

The success of sentiment-aware RL models in these experiments suggests broader implications for algorithmic trading. Traditional RL strategies often rely solely on historical price and volume data, which may limit their ability to capture sudden market shifts driven by news or macroeconomic events. By integrating qualitative insights through sentiment analysis, RL models can address this limitation, offering a more comprehensive approach to decision-making in dynamic financial markets.

Despite the promising results, several limitations should be acknowledged. First, the experiments were conducted using historical data, which does not fully replicate real-world market conditions, including slippage and transaction costs. While the models included small penalties for trade execution, future studies could incorporate more granular simulations to capture these factors more accurately. Additionally, the sentiment data used in these experiments relied on aggregated sentiment scores, which may oversimplify the complexities of financial news. Future research could explore more sophisticated sentiment extraction techniques, such as fine-grained sentiment analysis or context-aware models, to enhance the quality of qualitative inputs.

Another area for further exploration is the scalability of these approaches to larger and more diverse portfolios. While the RL models performed well on the LEXCX portfolio, extending this methodology to portfolios with greater sectoral or geographical diversity could provide additional insights into the generalizability of sentiment-aware trading strategies.

\section{Conclusion}
\label{sec:conclusion}
This study explored the application of reinforcement learning (RL) for financial trading in two distinct setups: single-stock trading with Apple Inc. (AAPL) and portfolio trading with the ING Corporate Leaders Trust Series B (LEXCX). By incorporating sentiment analysis derived from financial news, the RL models demonstrated significant improvements in trading performance compared to models that relied solely on quantitative stock data. This enhancement underscores the potential of integrating qualitative market signals into RL frameworks for financial decision-making.

The results of the experiments showed that both single-stock and portfolio RL models were effective in navigating complex market conditions and generating positive returns. In the single-stock experiment, the sentiment-enhanced RL model achieved higher net worth and cumulative profit compared to the baseline model, highlighting the added value of sentiment data in improving trading accuracy. Similarly, in the portfolio experiment, the sentiment-enhanced model outperformed the RL model without sentiment and the actual LEXCX portfolio, demonstrating its ability to leverage sentiment data to make adaptive and profitable trading decisions.

This study also provided a benchmark comparison against the actual LEXCX portfolio’s buy-and-hold performance. The RL agent, particularly with sentiment integration, consistently outperformed the passive investment strategy, showcasing its potential for dynamic portfolio management. These findings align with prior research, which emphasizes the importance of integrating diverse data sources, such as sentiment and market signals, to enhance the predictive capabilities of trading algorithms.

Despite these promising results, there are limitations to address in future work. First, the sentiment analysis relied on aggregated news data, which may not fully capture intraday market sentiment shifts. Incorporating real-time social media sentiment, such as Twitter or Reddit discussions, could provide a more comprehensive understanding of market psychology. Second, the models evaluated here used a fixed set of hyperparameters and market conditions, which may not generalize across diverse market regimes. Future studies could explore hyperparameter tuning, transfer learning, and model robustness across different market cycles.

In conclusion, this research highlights the transformative potential of combining RL with sentiment analysis in financial trading. By enabling agents to adapt to both quantitative and qualitative market signals, this approach provides a powerful tool for investors and institutions seeking to optimize trading strategies in increasingly complex and dynamic financial markets.

\break
\printbibliography

\end{document}